# Transfer-free Grown Bilayer Graphene Transistors for Digital Applications

Pia Juliane Wessely[1], Frank Wessely[1], Emrah Birinci[1], Bernadette Riedinger[2], Udo Schwalke[1]

[1]Institute for Semiconductor Technology and Nanoelectronics (ISTN), Technische Universität Darmstadt, Schlossgartenstrasse 8, 64289 Darmstadt, Germany

[2]Fraunhofer-Institut für Werkstoffmechanik, Wöhlerstrasse 11, 79108 Freiburg, Germany

**ABSTRACT**

We invented a novel method to fabricate graphene transistors on oxidized silicon wafers without the need to transfer graphene layers. By means of catalytic chemical vapor deposition (CCVD) the in-situ grown bilayer graphene transistors (BiLGFETs) are realized directly on oxidized silicon substrate, whereby the number of stacked graphene layers is determined by the selected CCVD process parameters, e.g. temperature and gas mixture. BiLGFETs exhibit ultra-high on/off-current ratios of $10^7$ at room temperature, exceeding previously reported values by several orders of magnitude. This will allow a simple and low-cost integration of graphene devices for digital nanoelectronic applications in a hybrid silicon CMOS environment for the first time.

## 1. INTRODUCTION

A monolayer of graphene consists of carbon atoms which are arranged in a quasi planar honeycomb lattice structure. It is a true 2D material which has been exfoliated from graphite for the first time by A. Geim and K. Novoselov [1] in 2004. However, size and position of the exfoliated graphene flakes varies randomly. In addition to this difficulty, adsorbed molecules like $O_2$ and $H_2O$ often accumulate at the interface between graphene and the substrate surface [2].
In order to avoid graphene transfer, epitaxial graphene grown on silicon carbide (SiC) has been proposed by de Heer and Berger [3]. Using this method fairly large graphene sheets can be realized on a SiC wafer without the need to transfer. However, when comparing with conventional silicon processing, this method is more expensive because of the SiC substrate. Furthermore, the process requires extraordinary high growth temperatures of about 1400°C and is therefore not compatible with conventional silicon CMOS processing. Although graphene growth on dielectrics has been demonstrated by various research groups [4-9], none of these research groups is growing graphene or graphite directly on thermally grown $SiO_2$ using conventional silicon substrates. Some of the materials are either extremely expensive or completely incompatible with CMOS process technology. In addition, graphene field-effect transistors (GFETs) cannot be turned off effectively due to the absence of a band gap. As a result the on/off current ratios typically observed are below 30 in monolayer graphene FETs (MoLGFETs) [10, 11] and around 100 in bilayer graphene FETs at room-temperature [12]. These on/off current values are insufficient for digital CMOS applications.

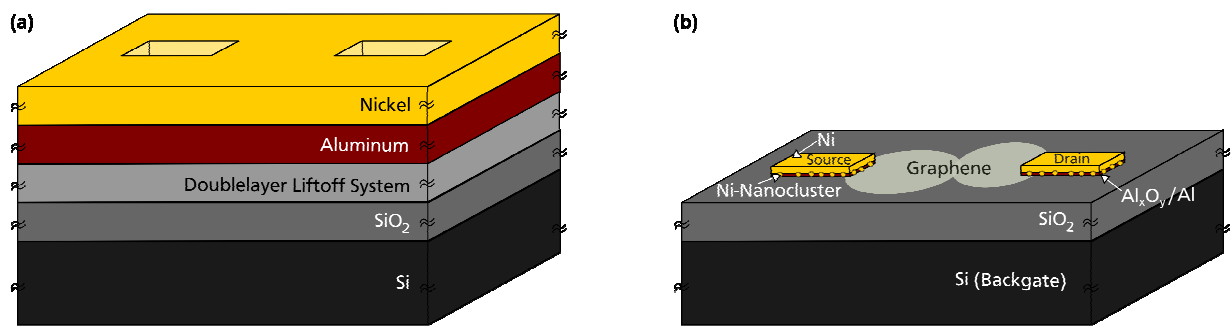

**Figure 1.** Schematic drawings illustrating the fabrication process of graphene transistors: (a) Double layer liftoff system used to pattern the catalyst areas. (b) Graphene FET structure produced by CCVD using an aluminum/nickel catalyst system. Note that the catalyst areas are simultaneously used as source and drain contacts.

We have developed a dedicated in-situ CVD-based growth method for graphene on oxidized silicon wafers in order to avoid the above mentioned drawbacks. First experimental evidence demonstrating the feasibility of this transfer-free graphene growth method has already been published in November 2009 [13, 14]. Furthermore, we observed extremely high on/off current ratios up to $10^7$ for bilayer graphene FETs [15].

## 2. EXPERIMENTAL

In preparation for CCVD a silicon wafer is oxidized in dry ambient at 1000°C for 120 min to obtain a 100nm thick $SiO_2$ film. Afterwards several lithography steps follow and a structured liftoff system remains on the wafer surface. Thin aluminum and nickel layers (5nm to 15 nm each) are evaporated over the whole substrate surface (cf. Fig 1a) and are structured via liftoff. By annealing the wafer at 800°C to 900°C for 3 to 15 minutes the aluminum transforms itself partially into aluminum-oxide-like insulator ($Al_xO_y$) while the nickel (Ni) layer dioxide surface (c.f. Fig. 1b), while the number of the stacked graphene layers depends on the adjusted process parameters in particular process time and temperature. The methane flow rates are typically in the range of 4 to 15 liters per minute while the methane can be diluted by hydrogen with a flow rate of 3 liters per minute at maximum at atmospheric pressure [10, 15, 16].

When using a suitable device layout, several hundred of large scale graphene FETs are fabricated simultaneously on one 2'' wafer and the graphene transistors are functional directly after the CCVD growth process, contacted directly via the catalytic source/drain (S/D) areas (cf. Fig. 1b) for electrical characterization. In addition to the grown graphene layers on the oxide surface further carbon deposits are present on top of the catalytic areas as evident from the enlargement shown in figure 2a. Detailed examination by means of scanning electron microscopy indicates that the spread of the CNTs and the additional carbon deposits occur exclusively on the top of catalyst source/drain (S/D) areas (cf. Fig. 2 magnification). The total processing time for the wafers within the CVD chamber (Applied Materials AMV 1200) is in the range of 30 to 60 minutes.

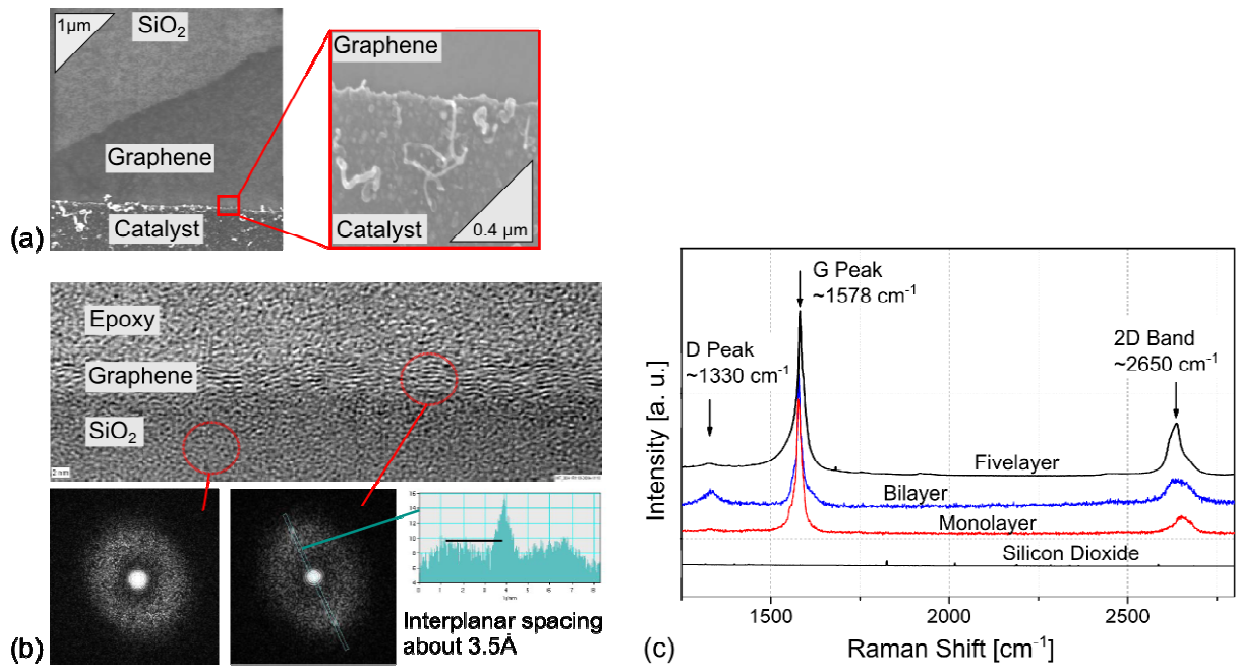

**Figure 2**. (a) HR scanning electron microscopy image of the probe at the catalyst graphene junction. (b) Structural transmission electron microscopy examination of a multilayer graphene sample on a silicon dioxide surface with Fourier analysis. (c) Raman spectra of in-situ CCVD grown fewlayer, bilayer and monolayer graphene measured within the channel region at room temperature. In addition, Raman measurements in approximately 50µm distance to the channel region were performed to establish a reference of the graphene-free silicon dioxide surface.

## 3. RESULTS AND DISCUSSION

A scanning electron microscopy image of a multilayer graphene sample at the catalyst to graphene transition can be seen in figure 2a. The surface of the catalyst area is rough and some carbon nanotubes growing from the nickel cluster are visible, as expected [13, 14]. Figure 2b shows the transmission electron microscopy (TEM) screening of the sample. Therefore the sample was mould in an epoxy resin. The Fourier analysis of structural transmission electron microscopy examination of the CCVD grown graphene multilayer on silicon dioxide shows an interplanar spacing of 3.5Å. This is a strong evidence for the existence of graphene grown by means of catalytic CVD [17]. Raman spectroscopy of graphene transistors is performed within the channel region in between the catalytic areas (cf. Fig. 1b) [18] and the results are shown in figure 2c. In the Raman spectra of graphene three main peaks can be determined: The D and G peak at 1338 $cm^{-1}$ and 1578 $cm^{-1}$ respectively, represent the graphitic $sp^2$-structure (G peak) and the defects in the graphene lattice, as holes and edges (D peak). The shape of the 2D band around 2650 $cm^{-1}$ is characteristic for the number of stacked graphene layers. Comparing figure 2c with Raman data from A. Ferrari [19] suggests the presence of monolayer and bilayer graphene in our sample. The characteristic Raman G and D peaks as well as the 2D band are located at similar Raman shift positions found by Ferrari and exhibit the expected shape [15]. However, a difference in the I(2D)/I(G) ratio especially for the monolayer graphene sample is observed. These differences indicate strong interactions of graphene with underlying silicon dioxide [20, 21] which are due to the in-situ growth at moderate temperatures. In addition, hydrogen doping is supposed to affect the I(2D)/I(G) ratio. Jaiswal et al. [22] investigated the Raman spectra of graphene sheets and nano ribbons with different degrees of hydrogenation by hydrogen plasma for different exposure times. Depending on the exposure time the I(2D)/I(G) ratio changed significantly. Since hydrogen is present during the CCVD growth single hydrogen atoms may substitute carbon atoms or attach to the border of the graphene layer which is known to affect respective Raman intensities [22].

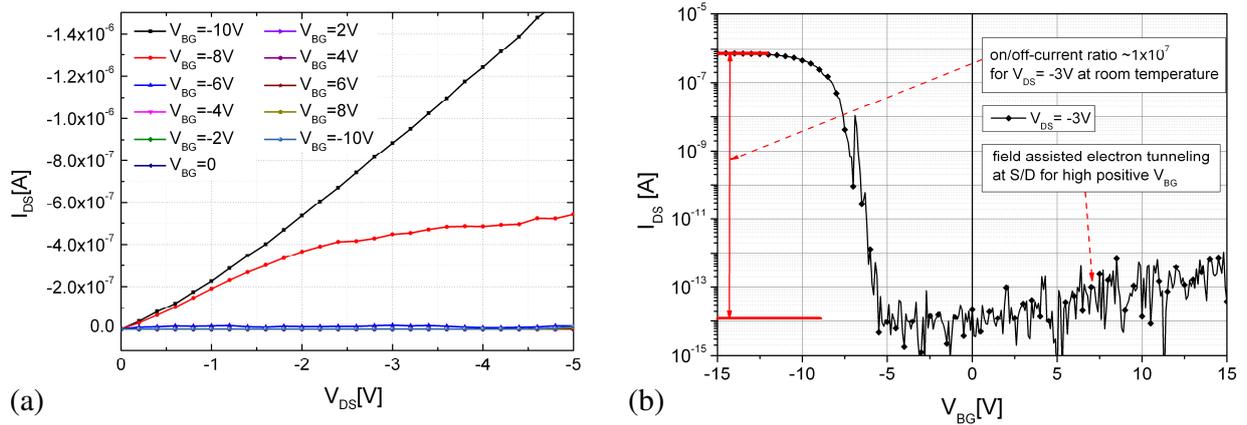

**Figure 3.** (a) Output characteristics of a bilayer graphene field effect transistor at different backgate voltages $V_{BG}$ while the drain/source voltage $V_{DS}$ is swept from -5V to 0V. (b) Input (i.e. current vs. gate-voltage) characteristics of a bilayer graphene transistor. The backgate voltage $V_{BG}$ is swept from -15V to 15V while a constant voltage $V_{DS}$ of -3V is applied between drain and source.

The electrical characterization of the graphene devices is performed using a Keithley SCS 4200 semiconductor analyzer. The metal catalyst areas are directly used as source and drain contacts (cf. Fig. 1b). However, for in-situ CCVD grown graphene FETs the maximum current is limited by the thin nickel conducting paths as well as the high contact resistance caused by some carbon deposits on top of the source drain regions [15,18]. Figure 3a shows the output characteristic of a BiLGFET. A constant backgate voltage $V_{BG}$ is applied while the drain/source voltage $V_{DS}$ is swept from -5V to 0V. The input characteristic of a bilayer graphene field effect transistor (BiLGFET) is shown in figure 3b. In-situ CCVD grown BiLGFETs exhibit an ultra-high on/off-current ratio of $1 \times 10^7$ which is to our knowledge the highest reported value for in-situ CCVD grown BiLGFETs today. The transfer characteristic is consistent with the output characteristic, displaying a unipolar, p-type device behavior. The required bandgap is partly realized by the applied backgate voltage causing an electrical field of low gate field strength of 1.5 MV/cm (i.e. 0.15V/nm) perpendicularly to the layer. In addition to this we assume a contribution of hydrogen doping as deduced from the Raman spectra as well as a high contribution of the interaction between the bilayer graphene and the substrate to the bandgap of bilayer graphene. Such intensive interactions may develop during the growth of the bilayer graphene on the silicon dioxide at moderate temperatures under well defined ambient conditions within a CVD chamber [153]. The band structure of BiLG is known to be sensitive to the lattice symmetry. If the two graphene

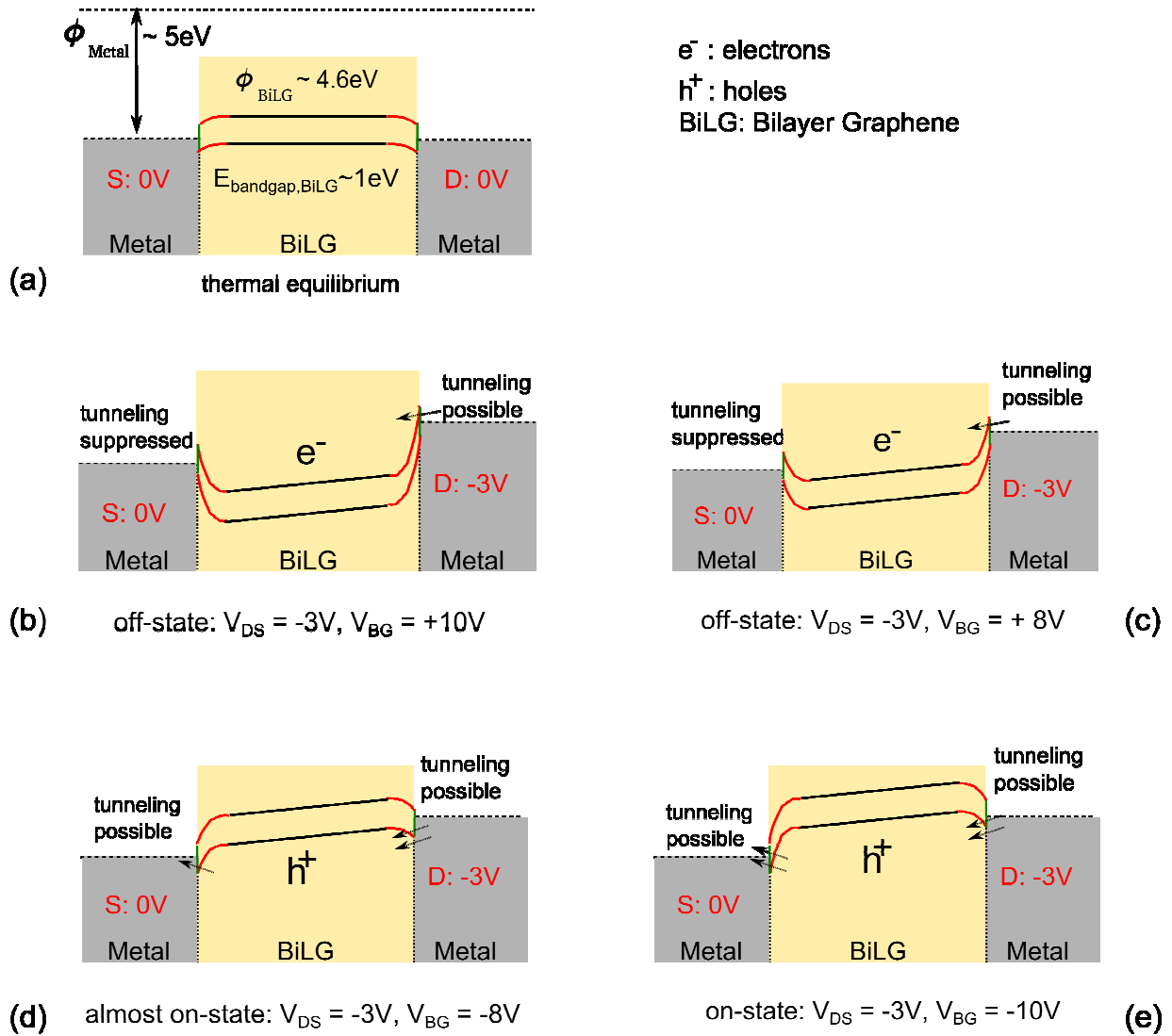

**Figure 4.** Schematic band diagrams regarding the output characteristic of a bilayer graphene field effect transistor (a) in thermal equilibrium. (b) Schematic band diagram for a BiLGFET when in the off-state, i.e. $V_{BG} = +10V$ at $V_{DS} = -3V$ (c) Schematic band diagram for a BiLGFET when in the off-state, i.e. $V_{BG} = +8V$ at $V_{DS} = -3V$ (d) Schematic band diagram for a BiLGFET while almost conducting in the on-state, i.e. $V_{BG} = -8V$ at $V_{DS} = -3V$. (e) Schematic band diagram for a BiLGFET fully conducting in the on-state at $V_{BG} = -10V$ and $V_{DS} = -3V$.

layers generate an asymmetric double layer, an energy gap between low-energy bands forms at the former Dirac crossing points [23]. Kong et al. [24] characterized the charge transfer of graphene to substrate by photoemission and inverse photoemission spectroscopy. The presented data demonstrate that a ~1 eV bandgap occurs for graphene on magnesium oxide MgO (111). The bandgap is due to graphene/substrate interfacial interactions [25]. Furthermore, the modulation of the Schottky-barrier at the Metal/BiLG contact by the applied backgate voltage in the range of $-2V < V_{BG} < 13V$ increases the off-state current due to field enhanced tunneling

(cf. Fig. 3b). This is may serve as an indication of a bandgap in the range of $0.6 < E_{bandgap, BiLG} < 1.0$ eV for the fabricated BiLG FETs.

Schematic band diagrams for a BiLGFET are shown in Figure 4 in order to explain the experimental observations of the output characteristic of a bilayer graphene field effect transistor. A metal work function of approximately 5eV for nickel is assumed as well as a band gap of approximately $E_{bandgap, BiLG} \sim 1.0$ eV for BiLG (cf. Fig. 4a). Previous studies [26, 27, 28, 29] reveal that the work function of graphene is in a similar range to that of graphite, ~4.6eV, [30] and depends sensitively on the number of layers [31, 32]. Figures 4b and 4c are showing the corresponding band diagrams of a BiLGFET in the off-state. In case of $V_{DS} = -3V$, $V_{BG} = +8V/+10V$ tunneling for electrons through the drain barrier is possible but tunneling through the source barrier is largely suppressed (cf. Fig. 4b,c). Simultaneously, hole conduction is also extremely unlikely. The on-state, i.e. when a voltage of $V_{DS} = -3V$ between drain and source is applied to the BiLGFET simultaneously with the back gate voltage of $V_{BG} = -10V$, can be seen in figure 4e. In this case the band bending is very severe enabling hole tunneling through the barriers at the drain and the source contact. $I_{DS}(V_{DS} = -3V, V_{BG} = -10V)$ is higher than $I_{DS}(V_{DS} = -3V, V_{BG} = -8V)$ as evident from figure 3a. The comparison of figure 4d and 4e shows that in case of $I_{DS}(V_{DS} = 3V, V_{BG} = -10V)$ hole conduction becomes more likely in case of a stronger band bending when compared to $I_{DS}(V_{DS} = -3V, V_{BG} = -8V)$. In fact, such a behavior is observed experimentally (cf. Fig. 3).

## 4. CONCLUSIONS

Various structural analyses have been performed on monolayer and bilayer graphene field effect transistors (MoLGFETs and BiLGFETs). The combination of structural transmission electron examination with Fourier analysis, Raman spectroscopy as well as extensive electrical characterization of graphene structures on silicon dioxide confirms the suitability of this in-situ CCVD growth process. MoLGFETs show the expected characteristic Dirac-point in the current-voltage characteristics and the typical low on/off-current ratios. In contrast, BiLGFETs exhibit ultra-high on/off-current ratios up to several $10^7$ at room temperature, exceeding previously reported values by several orders of magnitude. We explain the improved device characteristics by a combination of effects, in particular graphene-substrate interactions, hydrogen doping and Schottky-barrier effects at the source/drain contacts as well. With this novel fabrication method hundreds of BiLGFETs are realized simultaneously on one 2'' wafer by in-situ CCVD grown BiLG in a silicon CMOS compatible process. This will allow a simple and low-cost integration of graphene devices for nanoelectronic applications in a hybrid silicon CMOS environment for the first time.

**ACKNOWLEDGMENTS**

This research is part of the ELOGRAPH project within the ESF EuroGRAPHENE EUROCORES program and partially funded by the German Research Foundation (DFG, SCHW1173/7-1).